\documentstyle[emulateapj,apjfonts,psfig]{article}

\lefthead{J.~M.~Miller et al.}  
\righthead{SWIFT J1753.5$-$0127}

\received{Date}
\revised{Date}
\accepted{Date}

\journalid{vol}{date}
\articleid{1}{4}
\paperid{id}

\cpright{AAS}{1999}
\ccc{x}

\begin{document}

\title{A Prominent Accretion Disk in the Low--Hard State of the Black
Hole Candidate SWIFT J1753.5$-$0127}

\author{J.~M.~Miller\altaffilmark{1},
        J.~Homan\altaffilmark{2},
        G.~Miniutti\altaffilmark{3}}

\altaffiltext{1}{Department of Astronomy, University of Michigan, 500
Church Street, Ann Arbor, MI 48109, jonmm@umich.edu}
\altaffiltext{2}{Kavli Institute for Astrophysics and Space Research,
MIT, 77 Massachusetts Avenue, Cambridge, MA 02139}
\altaffiltext{3}{Institute of Astronomy, University of Cambridge,
Madingley Road, Cambridge CB3 OHA, UK}


\authoremail{jonmm@umich.edu}

\label{firstpage}

\begin{abstract}
We report on simultaneous {\it XMM-Newton} and {\it RXTE} observations
of the stellar-mass black hole candidate SWIFT J1753.5$-$0127.  The
source was observed in the ``low--hard'' state, during the decline of
a hard outburst.  The inner accretion disk is commonly assumed to be
radially truncated in the low--hard state, and it has been suggested
that this property may be tied the production of steady, compact jets.
Fits to the X-ray spectra of SWIFT J1753.5$-$0127 with a number of
simple models clearly reveal a cool ($kT \simeq 0.2$~keV) accretion
disk.  The disk component is required at more than the 8$\sigma$ level
of confidence.  Although estimates of inner disk radii based on
continuum spectroscopy are subject to considerable uncertainty, fits
with a number of models suggest that the disk is observed at or close
to the innermost stable circular orbit.  Recently, an observation of
GX~339$-$4 revealed a disk extending to the innermost stable circular
orbit at $L_{X}/L_{Edd} \simeq 0.05$; our results from SWIFT
J1753.5$-$0127 may extend this finding down to $L_{X}/L_{Edd} \simeq
0.003~ (d/8.5~{\rm kpc})^{2}~ ({\rm M}/10~{\rm M}_{\odot})$.  We
discuss our results within the context of low-luminosity accretion
flow models and disk--jet connections.

\end{abstract}

\section{Introduction}
Stellar mass black holes accreting at or below $10^{-2}~L_{Edd}$ are
found in the ``low--hard'' state (McClintock \& Remillard 2005).  This
state is typified by a hard, power-law X-ray spectrum ($\Gamma
\simeq$1.4--1.7), band-limited X-ray noise and high fractional
variability (sometimes low--frequency quasi-periodic oscillations, or
QPOs, are observed), and steady jet production that has been imaged in
at least one case (for reviews, see McClintock \& Remillard 2005 and
Fender 2005).  Recently, a deep observation of GX~339$-$4 was obtained
in the ``low--hard'' state (at $L_{X}/L_{Edd} \simeq 0.05$), using
{\it XMM-Newton} and {\it RXTE} (Miller et al.\ 2006).  The properties
of the thermal emission from the accretion disk and the width of a
relativistic Fe~K emission line reveal that the inner disk in
GX~339$-$4 is likely not radially truncated, in contrast to some
models for the inner accretion flow and jet production in this regime.
The moderate column density along the line of sight to GX~339$-$4 ($N_{H}
\simeq 4\times 10^{21}~{\rm cm}^{-2}$) enabled the study of disk
emission below 1--2~keV.

SWIFT J1753.5$-$0127 was discovered in hard X-rays with the {\it
Swift} Burst Alert Telescope (BAT) on 2005 May 30 (Palmer et al.\
2005).  The source was also clearly detected in soft X-rays with the
{\it Swift} X-ray Telescope (XRT) and in UV with the UV/Optical
Telescope (Morris et al.\ 2005, Still et al. 2005).  Optical
monitoring of SWIFT J1753.5$-$0127 revealed a double-peaked H$\alpha$
emission line, characteristic of the outer accretion disk in X-ray
binaries with a low-mass companion star (Torres et al. 2005).  At
present, the spectral type of the companion and the parameters of the
binary (e.g., orbital period, constituent masses, inclination) are not
known.

The hard power-law spectrum observed with the {\it Swift}/XRT (Morris
et al.\ 2005) and the detection of a 0.6~Hz QPO in pointed {\it RXTE}
observations are characteristic of the low--hard state (Morgan et al.\
2005).  SWIFT J1753.5$-$0127 was detected in radio observations made
with MERLIN at a flux density of 2.1 mJy at 1.7 GHz (Fender,
Garrington, \& Muxlow 2005), likely indicating compact jet activity.
Some black hole outbursts do not progress out of the low--hard state
(see, e.g., Brocksopp, Banyopadhyay, \& Fender 2004); public
lightcurves and hardness curves from the {\it RXTE}/ASM, and pointed
{\it RXTE} observations indicate that SWIFT~J1753.5$-$0127 has
remained in this state throughout its outburst.  The initial
part of the outburst followed a fast-rise, exponential-decay profile
over approximately 100 days, peaking at 0.2~Crab.  Thereafter, the
source has been persistently active at a low flux level (approximately
15 milliCrab).

To test whether or not the results from GX~339$-$4 hold at lower
factions of the Eddington limit, we requested a joint {\it XMM-Newton}
and {\it RXTE} observation of the stellar-mass black hole candidate
SWIFT J1753.5$-$0127.  Similar to GX~339$-$4, the moderate column density
to this source ($1.7\times 10^{21}~{\rm cm}^{-2}$, Dickey \& Lockman
1990) makes it an excellent target for such a study.  In the sections that
follow, we detail fits made to the {\it XMM-Newton} and {\it RXTE}
spectra of SWIFT J1753.5$-$0127 in the low--hard state, compare these
results to those obtained for GX~339$-$4, and discuss implications for
accretion onto stellar-mass black holes.

\section{Observations and Data Reduction}
{\it XMM-Newton} observed SWIFT J1753.5$-$0127 on 2006 March 24.  The
EPIC-pn exposure started at 16:00:31 UT, and lasted 42~ksec.  The
camera was operated in ``timing'' mode, and the ``medium'' optical
blocking filter was used for this observation.  We reduced the EPIC
data from the ODF level using SAS version 6.5.0, and the latest
calibration files.  An EPIC-pn event list was obtained by running the
processing task ``epproc''.  We then screened the pn data to exclude
bad events, bad pixels, and events registered too close to chip gaps
by requiring ``FLAG=0'' and ``PATTERN $\leq$ 4''.  Inspection of
lightcurves made from the 

\centerline{~\psfig{file=f1.ps,width=3.2in,angle=-90}~}
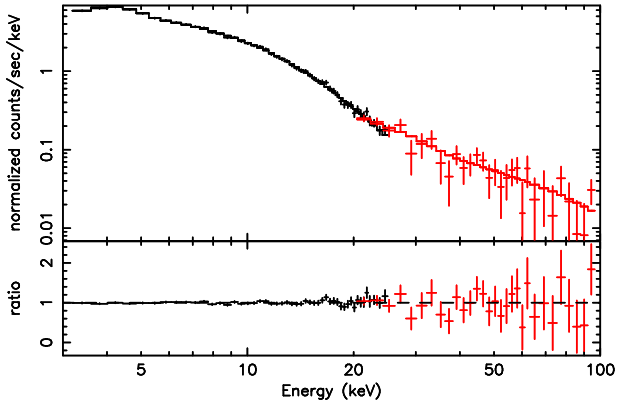
\figcaption[h]{\footnotesize The {\it RXTE} PCA (black) and HEXTE-B
(red) spectra of SWIFT J1753.5$-$0127 in the low--hard state are shown
above, fit with a simple absorbed power-law model.  This
phenomenological model is an acceptable fit to the {\it RXTE} spectra.}
\medskip

\noindent  data revealed no evidence of background flaring, and the
entire 42~ksec was used to generate EPIC-pn spectra.  Source and
background spectra were made in the standard way by grouping PI
channels 0--20479 by a factor of 5.  The source spectrum was extracted
between 20--56 in RAWX, and using the full RAWY range.  An adjacent
background region was extracted, and the spectra were properly
normalized.  Custom redistribution matrix files (rmfs) and ancillary
response files (arfs) for the spectra made using the SAS tasks
``rmfgen'' and ``arfgen''.  We also made MOS-1 and MOS-2 event lists
and spectra.  The MOS-1 camera was operated in ``timing'' mode, which
is seldom used.  The MOS-2 camera was operated in ``full-frame'' mode,
and therefore suffered photon pile-up.  While results from both MOS
cameras confirm the pn results we report below, we regard them as less
reliable and do not consider them in this work.

{\it RXTE} observed SWIFT J1753.5$-$0127 on 2006 March 24, starting at
17:22:24 UT.  This observation was reduced using the packages and
tools available in HEASOFT version 6.0.  After standard screening
(e.g., against SAA intervals), net PCA and HEXTE exposures of 2.3~ksec
and 0.8~ksec were obtained.  PCU-2 is the best calibrated PCU in the
{\it RXTE}/PCA, and so we extracted ``Standard2'' events (129 channels
covering 2--60~keV, obtained every 16 seconds) from this PCU.  Data
from all of the Xe gas layers in PCU-2 were combined.  Background
spectra were made using the FTOOL ``pcabackest'' using the latest
``bright source'' model.  An instrument response file was 
generated using the tool ``pcarsp''.  We added 0.6\%
systematic errors to the full PCU-2 spectrum using the ftool
``grppha'' (Miller et al.\ 2006).  We reduced ``archive'' mode
data from the HEXTE-B cluster; these data have a time resolution of 32~s
and cover the 10.0--250.0~keV band with 61 channels.  We extracted
source and background files, and generated an instrument response
file, using the standard procedures.

All spectra considered in this paper were grouped to require at least
10 counts per bin using the ftool ``grppha'' to
ensure valid results using $\chi^{2}$ statistical analysis.  The
spectra were analyzed using XSPEC version 11.3.2 (Arnaud \& Dorman
2000).  Fits made to the EPIC-pn spectrum were restricted to the
0.5-10.0~keV range by calibration uncertainies.  Similarly, fits to
the PCU-2 and HEXTE-B spectra were restricted to the 2.8--25.0~keV and
20.0--100.0~keV bands, respectively.  All of the error measurements
reported in this work are 90\% confidence errors, obtained by allowing
all fit parameters to vary simultaneously.

\centerline{~\psfig{file=f2.ps,width=3.2in,angle=-90}~}
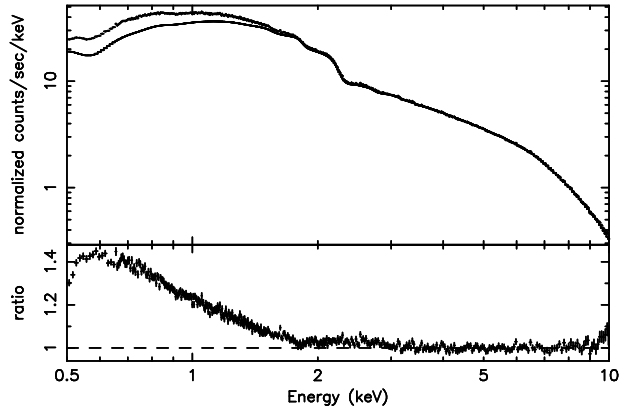
\figcaption[h]{\footnotesize The {\it XMM-Newton}/EPIC-pn spectrum of
SWIFT J1753.5$-$0127 is shown above.  The spectrum was fit with a
simple power-law model above 3~keV (to correspond to the {\it RXTE}
band), with the column density fixed at the best-fit value for
two-component disk plus power-law models (see Table 1).  The plot
above shows the soft disk excess that is clear when this model is
extended down to 0.5 keV.}
\medskip


\section{Analysis and Results}
Using standard FTOOLS, we made fast Fourier transforms of the {\it
RXTE} lightcurves of SWIFT J1753.5$-$0127 taken in PCA event modes.
The resulting power spectra show strong, band-limited variability that
is typical of the low--hard state in accreting black holes.  The rms
noise amplitude in the 0.01--100~Hz band is 30\%.  No QPOs were
detected in this observation.

Initial spectral fits were jointly made to the {\it RXTE} PCA and
HEXTE-B spectra with a simple absorbed power-law model.  A normalizing
constant was allowed to vary between the spectra, and the column
density was fixed at the expected value ($N_{H} = 1.7\times
10^{21}~{\rm cm}^{-2}$).  This fit gave a power-law index of $\Gamma =
1.65(2)$, and was in fact a formally acceptable result: $\chi^{2}/\nu
= 58.8/76$ (see Figure 1).  {\it RXTE} is not sensitive to variations
in low column densities, given that its effective lower energy
threshold is 3~keV.  We note that variations in the column density as
large as a factor of 2 only produced changes in the power-law index
within the error range quoted above, and also resulted in
statistically acceptable fits.  The power-law index measured with {\it
RXTE}, then, is a constraint robust against plausible variations in
the absorbing column.

We next made fits to the EPIC-pn spectrum, using a simple power-law
with the column density fixed at $1.7\times 10^{21}~{\rm cm}^{-2}$.
This model yielded an unacceptable fit ($\chi^{2}/\nu = 4855.2/1903$)
and strong residuals below 3~keV in the data/model ratio.  We next
allowed the column density to vary; this step yielded an improved but
unacceptable fit ($\chi^{2}/\nu = 3852.5/1901$).  The addition of a
disk component to this model yields significantly improved fits.  With
the addition of a disk component, the following parameters are
obtained $N_{H} = 2.3(1) \times 10^{21}~{\rm cm}^{-2}$, $kT =
0.22(1)$~keV, ${\rm Norm}_{disk} = 1200\pm 200$, $\Gamma = 1.66(1)$
(consistent with {\it RXTE}), and ${\rm Norm}_{pow} = 5.50(5)\times
10^{-2}$ ($\chi^{2}/\nu = 2227.0/1899$).  The soft excess
in this spectrum is shown in Figure 2, and the total
fit is shown in Figure 3.  The disk component is required at more than
the 8$\sigma$ level of confidence, as determined by an F test.  This
two-component model gives an unabsorbed flux of $3.9\times
10^{-10}~{\rm erg}~{\rm cm}^{-2}~{\rm s}^{-1}$ (0.5-10.0~keV) and a
luminosity of $3.4 \times 10^{36}~{\rm erg}~{\rm s}^{-1}~(d/8.5~{\rm
kpc})^{2}$ (or $L_{X}/L_{Edd} = 2.6\times 10^{-3}~(d/8.5~ {\rm
kpc})^{2}~({\rm M}/10~ {\rm M}_{\odot}$).  Of importance is the radius
of the inner disk; the normalization gives a radius of
$30(5)~(d/8.5~{\rm kpc})/{\rm cos}(i)$~km,

\centerline{~\psfig{file=f3.ps,width=3.2in,angle=-90}~}
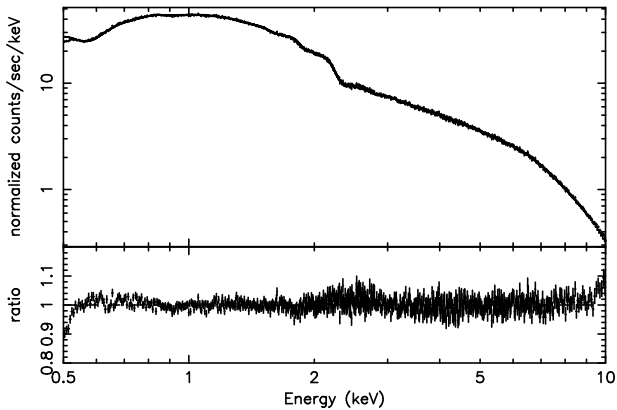
\figcaption[h]{\footnotesize The EPIC-pn spectrum of SWIFT
J1753.5$-$0127 is shown above, fit with a simple absobed disk
blackbody plus power-law model (see Table 1).  A number of variations
on this spectrum produce acceptable fits, and demonstrate that the
requirement of a disk which extends to the ISCO in
SWIFT~J1753.5$-$0127 is not stronly model-dependent.}
\medskip

\noindent or $2.0(3)~ ({\rm
M}/10 {\rm M}_{\odot})~ (d/8.5~{\rm kpc})/{\rm cos}^{1/2}(i)~
r_{g}$ (where $r_{g} = {\rm GM}/{\rm c}^{2}$).

For the disk blackbody radius given above, $r_{in} \leq 6 GM/c^{2}$
for $i \leq 83^{\circ}$ at 8.5~kpc, and for $i \leq 87^{\circ}$
(assuming $M = 10~M_{\odot}$) assuming $d = 2.9$kpc (see below).  At
such high inclinations, eclipses (not yet reported) would be expected.
For inclinations greater than $60^{\circ}$, $r \geq 6 GM/c^{2}$ would
nominally hold for $d \geq 18$~kpc assuming $M = 10~M_{\odot}$, or
for $d \geq 9$~kpc assuming $M = 5~M_{\odot}$.  However, it is likely
that SWIFT~J1753.5$-$0127 is closer than 8.5~kpc, making it even more
likely that the disk is close to the black hole.  Of the 18
dynamically-confirmed black hole binaries in the Milky Way, only one
may be as distant as 20~kpc (GS 1354$-$64, Casares et al.\ 2004) and
only one other is more distant than 8.5~kpc (GRS~1915$+$105,
e.g. Zdziarski 2005).  Torres et al. (2005) report $H=$14.8 on 11 July
2006.  On that day, an observation with the {\it RXTE}/PCA gives a flux of
$4.3\times 10^{-9}$~erg/s (2.8-12.0 keV), assuming a $\Gamma = 1.7$
power-law spectrum.  Russel et al.\ (2006) have recently found an
empirical relation between optical/IR luminosity and X-ray luminosity:
$L_{\rm OIR} = 10^{13.1} \times (L_{\rm X})^{0.61}$.  Given the
observed fluxes and a reddening of 0.19 for $N_{H} = 1.7\times
10^{21}~{\rm cm}^{-2}$, this relation implies a distance of 2.9~kpc to
SWIFT~J1753.5$-$0127.  Thus, the fit above implies that the accretion
disk in SWIFT~J1753.5$-$0127 is likely sitting at or near to the
innermost stable circular orbit, for a broad range of system
parameters.

Apart from uncertainties in distance, black hole mass, and
inclination, the choice of hard component model, disk model, and
radiative transfer through a disk atmosphere can all act to distort
the observed disk parameters.  For instance, Zimmerman et al.\ (2005)
show that enforcing a zero-torque inner boundary condition can drive
estimates of the inner radius values smaller by a factor of $\sim$2,
relative to the diskbb model.  Shimura \& Takahara (1995) suggest that
radiative transfer may act to give an inferred inner disk radius that
is too small by a factor of $\sim3$; Merloni, Fabian, \& Ross (2000)
suggest this factor may be as high as $\sim 9$ in some cases.  Torque
conditions and radiative transfer may partially cancel out, but
represent important additional uncertainties.  

In order to establish the nature of the disk as robustly as possible,
we jointly fit the EPIC-pn, PCU-2, and HEXTE-B spectra with several
continuum models.  For each model fit to the data, an overall
normalizing constant was allowed to float between the spectra.  We
employed three different disk models: ``diskbb'' (which assumes
maximal inner torque), ``ezdiskbb'' (which assumes zero inner torque),
and ``diskpn'' (which assumes zero inner torque, and a
pseudo-Newtonian inner potential).  For each disk model, we employed
three different models for the hard component: a simple power-law, a
hot optically--thin Comptonizing corona, and a cool optically--thick
Comptonizing corona.  The parameters obtained with these models are
listed in Table 1.  Each of these models suggests that the
disk in SWIFT J1753.5$-$0127 extended close to the innermost stable
circular orbit during our observations, for plausible ranges of source
distance, inner disk inclination, and black hole mass.  The
implication that the disk extends close to the black hole in SWIFT
J1753.5$-$0127 is not strongly model-dependent.  It is important to
note that the models with a low coronal electron temperature are
significantly worse than the others; they fail to produce the
hard flux required to fit the HEXTE-B spectrum.

If we assume that a disk temperature of $kT \simeq$ 1.0--2.0~keV is
typical for a stellar-mass black hole accreting at its Eddington limit
(e.g., McClintock \& Remillard 2005), the range of temperatures
implied by our fits is consistent with the theoretical expectation
that $T \propto \dot{M}^{1/4}$ in standard disks around black holes
(see, e.g., Frank, King, \& Raine 2002).  This theoretical consistency
indicates that the observed soft component is a simple accretion disk
continuum spectrum.

An independent observational constraint on the inner disk radius would
have been possible, if a strong relativistic Fe~K emission line like
that in GX~339$-$4 had been detected.  For a broad range of disk
inclinations, a relativistic ``Laor'' line consistent with moderate
black hole spin with an equivalent width of 60~eV cannot be ruled out.
This line strength would correspond to a reflection fraction of 0.3
for a disk with low ionization (George \& Fabian 1991), comparable to
the reflection constraints reported in stellar-mass black holes in
brighter phases of the low--hard state (e.g., Miller et al.\ 2006).  A
strong line (200--300~eV) like that detected in GX~339$-$4 (Miller et
al.\ 2006) is ruled-out in SWIFT~J1753.5$-$0127.

\section{Discussion and Conclusions}
Spectral and timing analysis of SWIFT~J1753.5$-$0127 reveal that the
source was in the low--hard state when observed simultaneously with
{\it XMM-Newton} and {\it RXTE} in March 2006.  Fits to the {\it
XMM-Newton}/EPIC-pn spectrum reveal clear evidence for an accretion
disk which likely extends close to the innermost stable circular orbit
around the black hole.  This result confirms prior results based on an
{\it XMM-Newton} and {\it RXTE} observations of the stellar-mass black
hole GX~339$-$4 (Miller et al.\ 2006) and {\it ASCA} observations of
Cygnus X-1 (Miller et al.\ 2006; also see Ebisawa et al.\ 1996 and
Balucinska \& Hasinger 1991) and likely extends those findings down an
order of magnitude in fractional Eddington luminosity, to
$L_{X}/L_{Edd} = 2.6\times 10^{-3}~(d/8.5~{\rm kpc})^{2}~({\rm
M}/10~{\rm M}_{\odot}$).  Whereas GX~339$-$4 was observed in a
low--hard state during the rising phase of its 2004 outburst, we
observed SWIFT~J1753.5$-$0127 during the decay phase of its 2005--2006
outburst.  It follows, then, that disks can remain close to the ISCO
in the low--hard state both in rising and decay phases of an outburst.

Advection--dominated accretion flow models (ADAFs; e.g. Esin,
McClintock, \& Narayan) predict that the inner disk should be radially
truncated in the low--hard state.  Indeed, in some black holes
observed at similar fractions of the Eddington limit (e.g., XTE
J1118$+$480, McClintock et al.\ 2001), and at lower luminosities, it
is possible that the disk may be radially recessed.  However, our
results suggest that this prediction may not hold universally in the
low--hard state.  A phenomenological model for jet production (Fender,
Belloni, \& Gallo 2004) suggests that a truncated disk may facilitate
jet production, in part because jet production does appear to be
largely quenched in ``high--soft'' states of stellar--mass black holes
(Fender 2005) wherein the inner disk is commonly thought to extend
close to the black hole.  Our observation of SWIFT J1753.5$-$0127
suggests that the jet production may not be enabled by a radially
truncated disk, nor quenched by a filled innermost accretion disk.
Other factors, perhaps including the mass accretion rate, or the
absence of a strong corona in high--soft states, may
inhibit jet production.

As previously noted, the spectral results obtained from this analysis
are not necessarily at odds with inferences drawn from X-ray timing
analyses of black holes in the low--hard state (Miller et al.\ 2006).
Low-frequency QPOs and noise components are not likely to be directly
related to Keplerian orbital frequencies at the ISCO.  In transitions
from the low--hard states to higher flux states, the frequency of
low-frequency QPOs are sometimes observed to saturate.  However, this
is again not necessarily indicative of the innermost stable circular
orbit, as such QPOs have been observed simultaneously with
high-frequency QPOs which are plausibly related to Keplerian orbits in
the inner disk (Remillard et a.\ 2002).

On theoretical grounds, it is all but impossible that a standard
optically-thick accretion disk can remain near the black hole in
the least luminous phases of the low--hard state as the source
approaches quiescence.  However, the fraction of the Eddington
luminosity at which the disk is actually truncated is yet to be
determined.  Exploring the regime near $L_X/L_{Edd} \simeq 10^{-4}$
will require an order of magnitude longer observation than this
43~ksec exposure.  Similarly, to obtain confirmation of the disk
radius at $L_X/L_{Edd} \simeq 10^{-3}$ via the width of an Fe~K
emission line, would likely also require a 400--500~ksec observation.

\vspace{0.1in}
We thank Norbert Schartel and the {\it XMM-Newton} staff, and Jean
Swank, Evan Smith, and the {\it RXTE} staff for executing these
observations.  We thank the anonymous referee for a helpful report.
GM thanks the UK PPARC for support.  This work has made use of the
tools and services available through HEASARC online service, which is
operated by GSFC for NASA.

\begin{table}[t]
\caption{Spectral Fit Parameters}
\begin{footnotesize}
\begin{center}
\begin{tabular}{llllllllll}
\multicolumn{2}{l}{Model/Parameter} & ${\rm N}_{\rm H} (10^{21}~{\rm cm}^{-2})$ & $kT$~(keV) & $r_{in}~(r_g)$ & $\Gamma$ & $kT_e$~(keV) & $\tau$ & Norm. & $\chi^{2}/\nu$ \\
\tableline

\multicolumn{2}{l}{diskbb+pow} & 2.3(1) & 0.23(1) & 1.9(2) & 1.66(1) & -- & -- & $5.50(5)\times 10^{-2}$ & $2286/1976$ \\

\multicolumn{2}{l}{ezdiskbb+pow} & 2.3(1) & 0.21(2) &  2.5(3) &  1.66(1) & -- & -- & $5.50(5)\times 10^{-2}$ &  $2290/1976$ \\

\multicolumn{2}{l}{diskpn+pow} &  2.3(1) & 0.21(1) & 6.0 & 1.66(1) & -- & -- &  $5.50(5)\times 10^{-2}$ &  $2290/1976$ \\

\tableline

\multicolumn{2}{l}{diskbb+comptt} & 1.6(1) & 0.18(1) & 2.4(4) & -- & 50.0 &  1.03(1) &  $3.1(1)\times 10^{-3}$ & $2158/1976$ \\

\multicolumn{2}{l}{ezdiskbb+comptt} & 1.7(1) & 0.17(1) & 3.0(4) & -- & 50.0 &  1.03(1) &  $3.1(1)\times 10^{-3}$ & $2158/1976$ \\

\multicolumn{2}{l}{diskpn+comptt} & 1.7(1) &  0.17(1) & 6.0 & -- & 50.0 & 1.03(1) & $3.0(1)\times 10^{-3}$ & $2159/1976$ \\

\tableline

\multicolumn{2}{l}{diskbb+comptt} & 1.7(1) & 0.16(1) & 2.2(5) & -- & 5.0 & 5.1(1) & $3.2(1)\times 10^{-2}$ & $2562/1976$ \\

\multicolumn{2}{l}{ezdiskbb+comptt} & 1.7(1) &  0.16(1) & 3.0(5) & -- & 5.0 & 5.1(1) & $3.2(1)\times 10^{-2}$ & $2566/1976$ \\

\multicolumn{2}{l}{diskpn+comptt} & 1.7(1) & 0.16(1) & 6.0 & -- & 5.0 & 5.1(1) & $3.2(1)\times 10^{-2}$ & $2566/1976$ \\

\tableline
\end{tabular}
\end{center} 
\tablecomments{The results of jointly fitting simple models to the
  {\it XMM-Newton}/EPIC-pn, and {\it RXTE} PCA and HEXTE spectra of
  SWIFT J1753.5$-$0127 are given above.  Each of the above models was
  modified by interstellar absorption using the ``phabs'' model.  An
  overall normalizing constant was allowed to float between the
  spectra.  The electron temperature and optical depth in the
  ``comptt'' model could not be constrained, so two extremes are
  examined above.  In both cases, the electron temperature of the
  corona is fixed at the quoted value.  Radii derived from the
  ``diskbb'' and ``ezdiskbb'' models are given in terms of
  $f^{2}~(d/8.5~{\rm kpc})~({\rm M}/10~{\rm M}_{\odot})/\sqrt
  cos(i)$, where $f$ is the spectral hardening factor and
  $i$ is the inner disk inclination (the ``diskbb'' model assumes
  $f=1$ and the ``ezdiskbb'' model assumes $f=1.7$).  Fits made with
  the ``diskpn'' disk model were made with the radius fixed at
  $6~r_g$, as suggested by the XSPEC notes on this model.}
\vspace{-1.0\baselineskip}
\end{footnotesize}
\end{table}

\end{document}